\documentclass[letter,scriptaddress,twocolumn, prl,showkeys,showpacs,superscriptaddress]{revtex4-1}

	\usepackage{amsmath}
	\usepackage{makeidx}
	\usepackage{amsfonts}
	\usepackage{mathrsfs}
	\usepackage{bbm}
	\usepackage{bm}
	\usepackage{esvect}
	\usepackage{graphicx, xcolor}  
	\usepackage{dcolumn}   
	\usepackage{bm}        
	\usepackage{amssymb}   
	\usepackage[ansinew]{inputenc} 
	\usepackage[usenames,dvipsnames]{pstricks}
	\usepackage{subfigure}
	\usepackage{epstopdf}
	\usepackage{pst-grad} 
	\usepackage{pst-plot} 
	\usepackage{multirow}
	\usepackage[colorlinks,hyperindex]{hyperref}
	\hypersetup
	{
		colorlinks,%
		citecolor=black,%
		linkcolor=black,%
		urlcolor=black,%
	}

\begin{document}

\title{Dynamical Phase Transitions Shape Biodiversity of a Migrating Population}
\title{Multiple Phase Transitions Shape Biodiversity of a Migrating Population}

\author{Casey O. Barkan}
\author{Shenshen Wang}
\email{shenshen@physics.ucla.edu}
\affiliation{Department of Physics and Astronomy, University of California, Los Angeles, Los Angeles, CA 90095}


\begin{abstract}

In a wide variety of natural systems, closely-related microbial strains coexist stably, resulting in high levels of fine-scale biodiversity.
However, the mechanisms that stabilize this coexistence are not fully understood. Spatial heterogeneity is one common stabilizing mechanism, but the rate at which organisms disperse throughout the heterogeneous environment may strongly impact the stabilizing effect that heterogeneity can provide. An intriguing example is the gut microbiome, where active mechanisms exist to control the movement of microbes and potentially maintain diversity. 
We investigate how biodiversity is affected by migration rate using a simple evolutionary model with heterogeneous selection pressure. We find that the biodiversity-migration rate relationship is shaped by multiple phase transitions, including a reentrant phase transition to coexistence. At each transition, an ecotype goes extinct and dynamics exhibit critical slowing down (CSD). CSD is encoded in the statistics of fluctuations due to demographic noise---this may provide an experimental means for detecting and altering impending extinction. 

\end{abstract}

\maketitle

Closely related microbial species, and even strains of the same species, can stably coexist in many natural systems~\cite{garud2019evolutionary,lieberman2016,oh2016temporal,tikhonov2015interpreting,acinas2004fine,rosen2018probing,rosen2015fine,conwill2022}, from phytoplankton living in the ocean or hot springs to bacteria residing in the human gut and skin. 
Understanding the stable coexistence of similar species has been a longstanding focus in ecology \cite{hutchinson1961paradox,chesson2000mechanisms,amarasekare2003competitive,ellner2019expanded}, and myriad mechanisms that may stabilize such coexistence have been examined~\cite{chesson2000mechanisms,pearce2020stabilization,roy2020complex}. Yet, the large extent of biodiversity at fine scales in nearby spatial locations remains a key puzzle~\cite{hart2017spatial,good2018effective,pearce2020stabilization}. Classic mechanisms that stabilize coexistence involve differences in resource consumption and asymmetric interactions between species~\cite{chesson2000mechanisms}. However, these mechanisms appear inadequate to explain fine-scale diversity of closely related strains having similar resource requirements and symmetric interactions~\cite{pearce2020stabilization}. 

It has recently been proposed that source-sink dynamics in spatially extended systems~\cite{dias1996sources,hermsen2010sources, hermsen2012rapidity}, chaotic coexistence or asynchrony~\cite{pearce2020stabilization}, and feedback from endogenous fluctuations in highly diverse microbial systems~\cite{roy2020complex} can sustain similar species for very long times. Common to these mechanisms is spatial loci connected by migration fluxes. Spatial heterogeneity can provide local niches to which different species or strains -- or more generally, ecotypes -- can specialize~\cite{amarasekare2003competitive}. Dispersal of organisms out of their niches may result in spatial coexistence of related ecotypes. Intuitively, one would expect a non-monotonic dependence of biodiversity on migration rate: Although elevated diversity relies on migration, excessive migration can reduce diversity because native strains are subject to increased competition from invaders~\cite{hart2017spatial}. 



But the reality is more complex. In a variety of systems where biodiversity has a strong impact on human health, evolution takes place on overlapping timescales as ecological dynamics. Significant examples include the gut microbiome~\cite{zhao2019,wolff2021ecological,garud2019evolutionary} and the adaptive immune repertoire~\cite{shai2020, sheng2021}. In addition, these systems appear to actively regulate the movement of evolving entities (microbes or antibody-producing cells) across structured environments (gut compartments or lymph nodes)~\cite{arnoldini2018bacterial, laichalk2002, bende2007}. An important challenge is, therefore, to understand in quantitative terms the influence of migration on biodiversity under heterogeneous selection pressure over long timescales. Such understanding could allow control of biodiversity by modulating migration or heterogeneity. 


In this article, we use a simple evolutionary model to show that the relationship between biodiversity and migration rate can be surprisingly complex. Our model, similar to that studied in~\cite{waclaw2010dynamical}, involves a heterogeneous environment comprising two habitats coupled by one-way migration. We find that the biodiversity-migration rate relationship is shaped by multiple phase transitions at which an ecotype goes extinct and the dynamics exhibit critical slowing down (i.e. the timescale of relaxation toward steady state diverges). Resource availability and selection profile in constituent habitats determine the number and location of phase transitions that can occur, which, in turn, determines how migration affects biodiversity. 
In particular, there is one regime where biodiversity is high over a broad range of migration rates. In a different regime, biodiversity reaches a maximum at a modest migration rate, falls to a lower level at intermediate rates, then, after crossing a re-entrant phase transition, returns to the same maximum under high migration.

Critical slowing down (CSD) near ecological tipping points has been widely studied in ecology, often considered a warning signal for impending ecological collapse~\cite{wissel1984universal,drake2010early,dai2012generic, scheffer2015generic}. It was recently argued that evolutionary changes should be incorporated into this picture \cite{dakos2019ecosystem}. 
Here, we begin with an evolutionary model where ecotypes emerge through evolution, instead of being pre-defined. 
With sufficient heterogeneity, distinct ecotypes evolve and compete for resources. At certain critical migration rates, evolutionary driving forces -- mutation, selection, and migration -- reach a balance, resulting in a vanishing restoring force to the steady state and hence critical slowing down. At these transitions, dynamics is driven solely by resource competition between ecotypes. 


\textit{Model} --- We model a population of asexual organisms, each characterized by a genotype $i\in\{1,...,M\}$, that inhabit two habitats. Habitats 1 and 2 have fitness landscapes $\vec\phi_1$ and $\vec\phi_2$, respectively, whose components, $\phi_{l,i}$, specify the fitness of genotype $i$ in habitat $l$. Organisms migrate from habitat 1 to habitat 2 at rate $k$. This one-way migration mimics the directed movement of evolving entities guided by a fluid flow or a chemical gradient. Mutations occur at a rate $\gamma$. Each habitat has a resource constraint of strength $\rho_l$, modelled by a logistic term that limits the growth of all genotypes equally (i.e., no imposed niches from interactions). As a result, a stable coexistence of distinct ecotypes, if any, would be due to heterogeneity.  We are working in the strong-selection weak-mutation regime, where crossing of fitness valleys due to mutation is negligible. The genotype abundances in each  habitat are specified by a vector $\vec n_l$ ($l=1,2$). 
Before introducing demographic noise, we first consider population dynamics under a deterministic model:
\begin{align}
    \frac{d}{dt}\vec n_1 = (\hat{V}_1-\rho_1 n_1^\textrm{tot})\vec{n}_1-k\vec n_1 \label{eq:EOM1}\\
    \frac{d}{dt}\vec n_2 = (\hat{V}_2-\rho_2 n_2^\textrm{tot})\vec{n}_2 +k\vec n_1 \label{eq:EOM2}
\end{align}
where $n_l^\textrm{tot}=\sum_{i=1}^M n_{l,i}$ is the total population of habitat $l$. $\hat V_1$ and $\hat V_2$ are $M\times M$ evolution matrices with elements
\begin{equation}\label{eq:V}
    \hat V_{l,ij} \equiv \delta_{ij} \phi_{l,i} +\gamma \Big( \hat A_{ij} - \delta_{ij}\sum_{k=1}^M \hat A_{ki} \Big)
\end{equation} where $\delta_{ij}$ is the Kronecker delta. The $\gamma$-dependent term describes mutation-induced genotypic turnover. The adjacency matrix $\hat A$ specifies the connectivity of genotype space: we assume that $\hat A_{ij}$ equals 1 if genotypes $i$ and $j$ differ by a single mutation and equals 0 otherwise. Importantly, our results are valid for any symmetric genotype adjacency matrix~\cite{SM}. In 1D (Figs. 1--3), genotype $i$ can only mutate to $i-1$ or $i+1$. This choice allows for easy visualization of the dynamics and steady states but is not necessary for our conclusions to be valid. 



We will first derive the steady-state populations at varying migration rates, and then examine the effect of fluctuations due to demographic noise. This approach is motivated by recent studies showing that populations of bacterial strains in the human gut often fluctuate around a stable value over the span of years~\cite{faith2013long}, and that the statistics of fluctuations are accurately described by a stochastic logistic model~\cite{wolff2021ecological}.


The specific questions we address within our model are: (1) What condition must spatial heterogeneity satisfy for multiple ecotypes utilizing the same resource to persist stably? (2) How does biodiversity depend on migration rate in heterogeneous environments? (3) What biological mechanism gives rise to the phase transitions that shape the biodiversity-migration rate relationship?

\textit{Coexistence of distinct ecotypes under sufficient heterogeneity} --- Evolutionary dynamics described by Eqs.~\ref{eq:EOM1} and \ref{eq:EOM2} lead to populations that depend on the dissimilarity between the fitness landscapes. Sufficient heterogeneity allows for coexistence of genetically distinct ecotypes, contributing to biodiversity and giving rise to phase transitions. We will analyze the steady states and simulate the dynamics (see supplemental movies~\cite{SM}).

In habitat 1, from any initial condition, $\vec n_1$ tends towards a steady-state $\vec n^*_1$ (asterisks denoting steady-state quantities hereafter) given by
\begin{equation}\label{eq:n1}
\vec{n}_1^* = \begin{cases} 
      \big(1-\frac{k}{\lambda_1}\big)K_1\vec{\psi}_1 & k\leq \lambda_1 \\
      0 & k > \lambda_1
\end{cases}
\end{equation}
where $\vec\psi_1$ -- the eigenvector of $\hat{V}_1$ associated with the largest eigenvalue $\lambda_1$ -- represents a normalized distribution of genotype frequencies (i.e. $\sum_i\psi_{1,i}=1$). $K_1\equiv\lambda_1/\rho_1$ is the carrying capacity of habitat 1. Migration discounts the total population to below the native value at $k=0$. Fig.~1a shows $\vec\psi_1$ (proportional to $\vec n_1^*$) for two example systems.
 
An essential observation is that $\vec\psi_1$ defines a distribution of genotypes that is exponentially localized around the most-fit region of the fitness landscape $\vec\phi_1$ (see SM \cite{SM}). In other words, $\vec\psi_1$ represents a localized cluster in genotype space. 
We identify such clusters as \textit{ecotypes}. A population with distribution $\vec\psi_1$ (which we call ecotype 1) represents the \textit{native ecotype} of habitat 1, in that it fills the spatially-localized niche formed by this habitat. The corresponding eigenvalue $\lambda_1$ measures the intrinsic growth rate of ecotype 1, i.e. the growth rate at low abundance when resource limitation poses no constraint. Similarly, habitat 2 has a carrying capacity $K_2=\lambda_2/\rho_2$ and native ecotype (ecotype 2) described by $\vec\psi_2$, where $\lambda_2$ is the largest eigenvalue of $\hat{V}_2$ and $\vec\psi_2$ its eigenvector.

\begin{figure}[b]
\includegraphics[width=8.5cm]{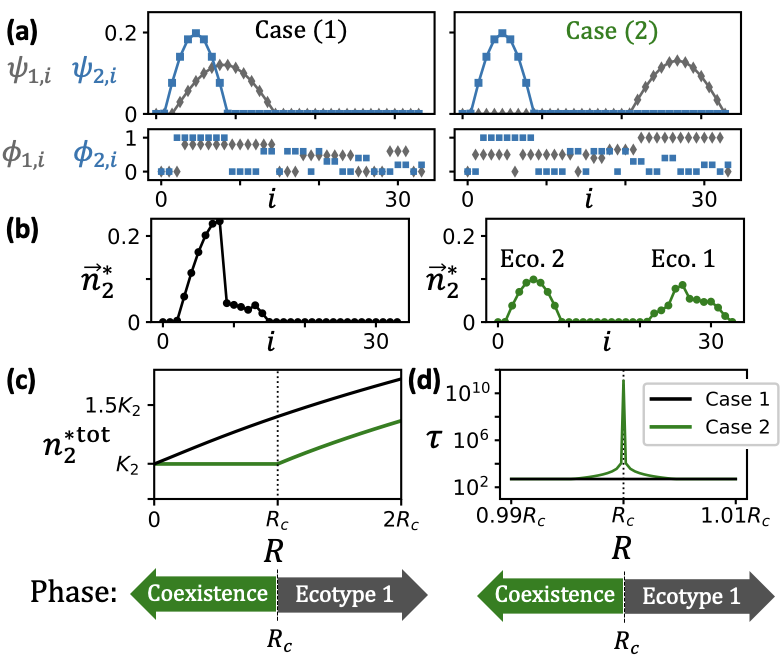}
\caption{\label{fig:1} 
\textbf{Coexistence of genetically-distinct ecotypes leads to phase transitions.} \textbf{(a)} In case 1 (left), native ecotypes $\vec\psi_1$ (grey) and $\vec\psi_2$ (blue) share common genotypes, but in case 2 (right) they are genetically distinct. Lower panels: fitness landscapes $\phi_{1,i}$ (grey diamonds) and $\phi_{2,i}$ (blue squares). \textbf{(b)} Steady-state population in habitat 2 with $R=R_c/2\approx0.41$ ($k\approx0.10$) and $\hat{R}=\vec\psi_1$ in two cases. In case (1), habitat 2 supports a single ecotype. The genotypes shared by $\vec\psi_1$ and $\vec\psi_2$ have the highest abundance. In case (2), habitat 2 supports coexistence of ecotype 1 (genotypes 22 to 33) and ecotype 2 (genotypes 2 to 10). \textbf{(c)} Habitat 2's total population in case 1 (black) and case 2 (green). In case (2), habitat 2's total population remains constant at the carrying capacity $K_2$ until $R$ reaches $R_c$. \textbf{(d)} Relaxation time $\tau$. $\tau\to\infty$ at $R=R_c$ in case (2), but not in case (1). Phases of \textit{Coexistence} ($R<R_c$) and \textit{ecotype 1} ($R>R_c$) are indicated. $\rho_1=0.2$, $\rho_2=1$, $\gamma=0.01$. }
\end{figure}


The behavior of habitat 2 is more interesting because of the influx of migrants from habitat 1. It is useful to consider the migrant flux $\vec R=k\vec n_1$, which gives the number of organisms of each genotype that enter habitat 2 per unit time. $\vec R$ can be decomposed as $\vec R=R(k)\hat{R}$ where $\hat R$ (normalized so that $\sum_j\hat{R}_j=1$) indicates the genotype composition and $R(k)$ the flux magnitude at migration rate $k$. Eq.~\ref{eq:n1} implies that
\begin{equation}
    R(k) = \begin{cases} 
      k(\lambda_1-k)/\rho_1 & k\leq \lambda_1 \\
      0 & k > \lambda_1
\end{cases}
\end{equation}
when the population of habitat 1 is at steady state (Fig. 2a). While flux $R$ must increase with migration rate $k$ for low $k$, habitat 1's population is depleted as $k$ is increased, so that $R(k)$ \textit{decreases} with $k$ for $k>\lambda_1/2$. We will show that this non-monotonic dependence of $R$ on $k$ gives rise to the re-entrant phase transition.

The effect of influx magnitude $R$ on the receiver population has an intriguing dependence on environmental contrast between the two habitats (dissimilarity between $\vec{\phi}_1$ and $\vec{\phi}_2$). Consider two cases -- \textbf{Case~(1)}: the two habitats have sufficiently similar selection pressure so that their native ecotypes (represented by $\vec\psi_1$ and $\vec\psi_2$) share common genotypes, implying that $\vec\psi_1\cdot\vec\psi_2>0$ (Fig. 1a left panel). In other words, $\vec\psi_1$ and $\vec\psi_2$ form a single cluster in genotype space. As we will show, in this case, the system does not undergo a transition that requires competition between distinct ecotypes. \textbf{Case~(2)}: the two habitats have sufficiently different selection pressure so that their native ecotypes have no genotype in common, implying that $\vec\psi_1\cdot\vec\psi_2=0$ (Fig. 1a right panel). In this case, $\vec\psi_1$ and $\vec\psi_2$ represent genetically distinct ecotypes; their competition for resources underlies the slow relaxation near a re-entrant transition.
Since genotype frequencies are continuous variables, $\vec\psi_1\cdot\vec\psi_2=0$ is never satisfied exactly in our model; nevertheless, it is often an excellent approximation for biologically realistic fitness landscapes~\cite{SM}.

In case (1), habitat 2's steady-state population can be obtained by solving $\frac{d}{dt}\vec n^*_2=0$, which yields $\vec n_2^* = (\rho_2 n_2^{*\textrm{tot}}-\hat{V}_2)^{-1} \vec{R}$ (Fig. 1b left panel). As one would expect, habitat 2's total population $n_2^{*\textrm{tot}}$ increases smoothly with $R$ (Fig. 1c black curve) and, as $R\rightarrow \infty$, $n_2^{*\textrm{tot}}\rightarrow \sqrt{R/\rho_2}$. 
Although the exact genetic makeup of $\vec n_2^*$ varies with $R$, certain genotypes are present for all values of $R$.

In case (2), as $R$ increases, surprisingly, $n_2^{*\textrm{tot}}$ remains constant at the value $K_2$ until $R$ surpasses a critical value $R_c \equiv \eta\lambda_2K_2/4$, where $\eta$ is a constant~\cite{SM} (Fig. 1c green curve). 
As a result, the expression for $\vec n_2^*$ given for case (1) is inapplicable, because $(\rho_2 n_2^{*\textrm{tot}}-\hat{V}_2)$ is no longer invertible. Instead,
\begin{equation}\label{eq:nBelow}
    \vec{n}_2^* = \bigg(1-\frac{R}{R_c}\bigg)K_2\vec\psi_2 + \frac{R}{R_c}\vec{J}
\end{equation}
for $R\leq R_c$ (Fig. 1b right panel). Here, $\vec{J}=(\lambda_2-\hat{V}_2)^+\vec R_c$ with the symbol $^+$ denoting pseudoinverse, and $\vec R_c=\hat R R_c$. A key intuition from this result is that $(\lambda_2-\hat{V}_2)^+$ is a linear filter that accounts for the selection pressure faced by immigrant organisms in habitat 2. $\vec J$ is the result of filtering $\vec R_c$, the migrating population at critical flux, and represents ecotype 1 in habitat 2. 
Due to the filter, $\vec J$ is modified from its native distribution $\vec\psi_1$. However, $\vec J$ is localized around the same region of genotype space as $\vec\psi_1$ and, like $\vec\psi_1$, shares no common genotypes with $\vec\psi_2$ (comparing the right panel of Figs. 1a and 1b). Eq.~\ref{eq:nBelow} shows that ecotypes 1 and 2 coexist in habitat 2 for $R<R_c$. As $R$ is increased, ecotype 1 steadily displaces ecotype 2, and when $R\geq R_c$, ecotype 2 goes extinct and ecotype 1 achieves competitive exclusion. Quantitatively, $\vec n_2^*=(\rho_2n_2^{*\textrm{tot}}-\hat{V}_2)^{-1} \vec{R}$ for $R> R_c$. 

Fig. 1d shows that $\tau$, the timescale of relaxation toward steady-state (discussed in detail below), is sharply peaked around $R_c$ in case (2). In contrast, $\tau$ in case (1) remains constant. The peak in $\tau$ arises due to interaction between genetically distinct ecotypes, and therefore, only systems with sufficient heterogeneity to fall under case (2) will exhibit this phenomenon. 
Hence, spatial heterogeneity able to support distinct ecotypes using the same resource will significantly alter steady state populations and dynamics near steady state.

The concurrence of ecotypes 1 and 2 in the same spatial location is an example of \textit{adaptive} diversity, characterized by coexistence of genotypes that fill distinct ecological niches. In our model, each habitat creates a niche, and migration causes ecotypes of different niches to coexist spatially. 
In contrast, as shown by Eq.~\ref{eq:nBelow}, \textit{neutral} diversity within each ecotype is independent of migration flux (distributions $\vec\psi_2$ and $\vec J$ are independent of $R$). Hence, changing migration rate alters adaptive diversity while leaving neutral diversity unchanged. Such adaptive diversity exists only in a case (2) system. Thus, to the question of ``how much" heterogeneity is needed to support adaptive diversity, our answer is: sufficient heterogeneity to satisfy case (2).

\textit{Biodiversity--Migration rate relationship} --- As seen above, migration has a strong impact on genotypic diversity of the steady-state populations. In particular, the relationship between biodiversity and migration flux appears to be shaped by the phase transition that occurs at $R=R_c$. To quantify this relationship, we measure biodiversity using the second order Renyi entropy:
\begin{equation}\label{renyi}
    H[\vec{n}] = -\log \sum_{i=1}^M \bigg(\frac{n_i}{\sum_j n_j}\bigg)^2.
\end{equation}
Intuitively, $H[\vec{n}]$ measures the (negative) log likelihood of finding two randomly picked individuals to have an identical genotype.
Since we are primarily interested in how spatial heterogeneity supports adaptive diversity, we focus on case (2) hereafter. Interestingly, for case (2) systems, the flux that maximizes biodiversity is found to be: $R_H = R_c/(1+e^{ H[\vec\psi_2] - H[\vec J]})$, which depends only on the difference in neutral diversity between the two ecotypes~\cite{SM}. This establishes that there is a non-monotonic dependence of biodiversity on migrant flux, with $H$ reaching a unique maximum at $R_H$. However, the dependence of biodiversity on $k$ is more complex.

To understand the effect of migration \textit{rate} $k$, as opposed to \textit{flux} $R$, we re-examine the steady state populations in terms of $k$. Eq.~\ref{eq:nBelow} shows that ecotype 2 will go extinct if $R$ is held above $R_c$; but does habitat 1 have the capacity to output such a flux of migrants? Recall that $R=k(\lambda_1-k)/\rho_1$ which has a maximum $R_\textrm{max}=\lambda_1K_1/4$. If $R_\textrm{max}<R_c$ then habitat 1 lacks the capacity to allow ecotype 1 to invade habitat 2 and wipe-out its native ecotype (Fig.~2 left column). However, if $R_\textrm{max}>R_c$ (Fig.~2 right column), then there are two critical migration rates, $k_+$ and $k_-$, at which $R=R_c$, given by
\begin{equation}\label{eq:kpm}
    k_{\pm} = \frac{\lambda_1}{2} \Big(1 \pm \sqrt{1-\eta\lambda_2 K_2/\lambda_1K_1} \Big).
\end{equation}
The condition $R_\textrm{max}>R_c$ can be re-stated as $\lambda_1K_1>\eta\lambda_2K_2$. Intuitively, this reflects the fact that habitat 1's carrying capacity $K_1$ and intrinsic growth rate $\lambda_1$ must be sufficiently large relative to habitat 2's in order for habitat 1 to produce sufficient migrant flux to eliminate ecotype 2. When this condition is met, a migration rate between $k_-$ and $k_+$ will drive ecotype 2 to extinction and ecotype 1 will achieve competitive exclusion in the two-habitat system. 

\begin{figure}[b]
\includegraphics[width=8.5cm]{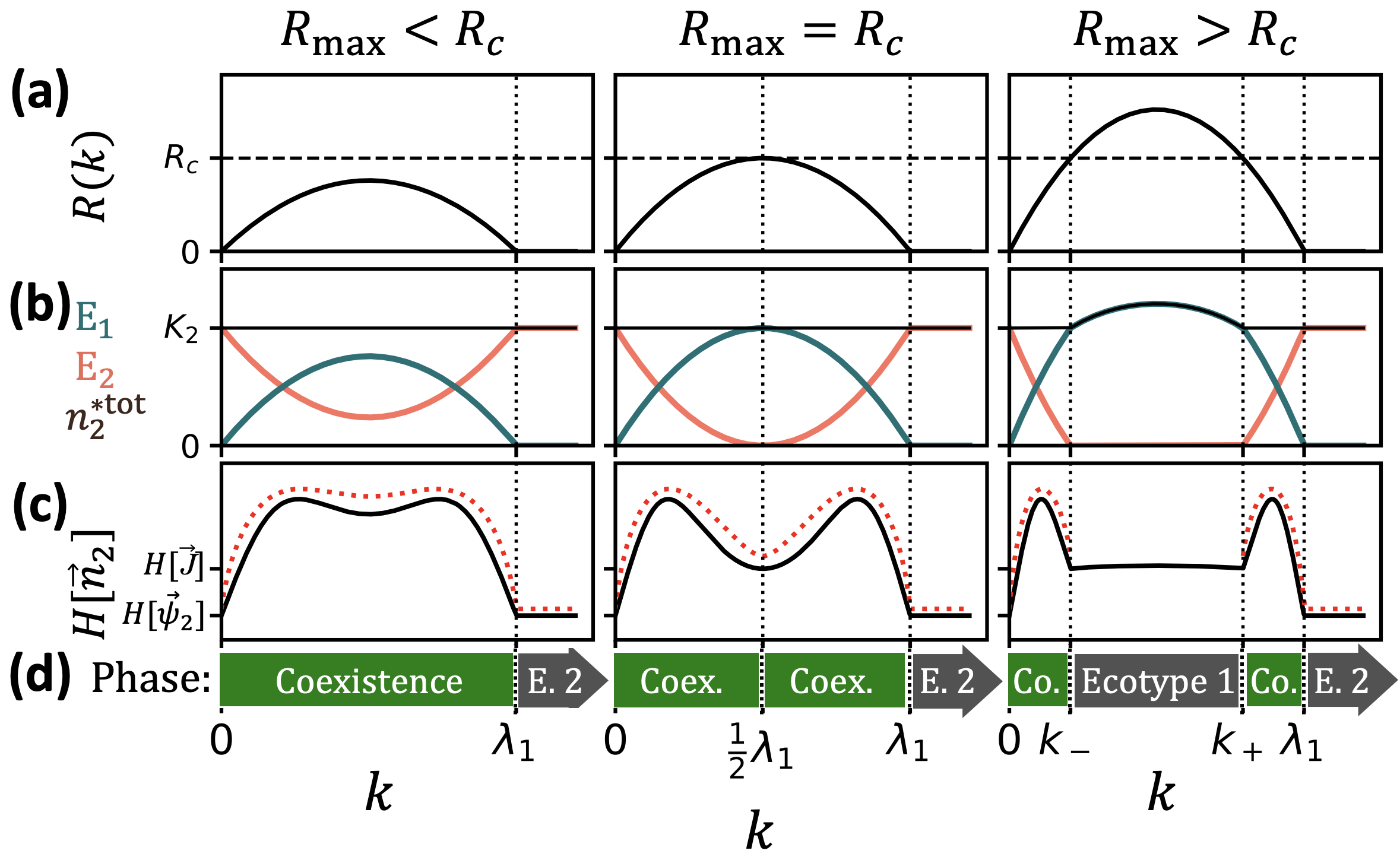}
\caption{\label{fig:1} 
\textbf{Phase transitions shape the relationship between biodiversity and migration rate.} \textbf{(a)} Migrant flux $R(k)$ as a function of $k$ for a system with $R_\textrm{max}<R_c$ (left), $R_\textrm{max}=R_c$ (middle), and $R_\textrm{max}>R_c$ (right). \textbf{(b)} Steady state total population of ecotype 1, $E_1$ (teal) and ecotype 2, $E_2$ (orange) in habitat 2, and habitat 2's total population $n_2^{*\textrm{tot}}$ (black). \textbf{(c)} Biodiversity of habitat 2, $H[\vec n_2]$, defined by the second-order Renyi entropy (solid black) and Shannon entropy (dotted red). \textbf{(d)} Phases in different regimes. Diagrams indicate the coexistence phase and phases with only ecotype 1 or ecotype 2 (E.~2). The regime of $R_\textrm{max}>R_c$ is characterized by the presence of ecotype 1 phase and the reentrance of coexistence phase at intermediate $k$ values, both of which are absent in other regimes. All panels show the same setting as in Fig. 1 case (2), except with $\rho_1=0.4$ for $R_\textrm{max}<R_c$ (left), $\rho_1=0.304$ for $R_\textrm{max}=R_c$ (middle), and $\rho_1=0.2$ for $R_\textrm{max}>R_c$ (right).}
\end{figure}

From the steady-state behavior, we can characterize the system by defining three distinct phases (Fig.~2d): the \textit{coexistence} phase where ecotypes 1 and 2 stably coexist, and two single-ecotype phases, where only ecotype 1 or ecotype 2 persists. 
If $R_\textrm{max}<R_c$ (Fig.~2 left), then the coexistence phase occurs over the wide range $0<k<\lambda_1$. On the other hand, if $R_\textrm{max}>R_c$ (Fig.~2 right), then the coexistence phase first occurs within the narrower range $0< k <k_-$. For $k_-<k<k_+$, the system exhibits the ecotype 1 phase (only ecotype 1 persists stably), then for $k_+<k<\lambda_1$ the system re-enters the coexistence phase through the re-entrant transition at $k=k_+$. In the singular case that $R_\textrm{max}=R_c$ (Fig.~2 middle), coexistence is stable over the entire range $0<k<\lambda_1$, except at a single point $k=\frac{1}{2}\lambda_1$ where ecotype 2 vanishes. In all three cases (Fig.~2 left, middle, and right), the ecotype 2 phase occurs for $k\geq \lambda_1$. As expected, biodiversity (Fig.~2c solid black curves) is generally high in the coexistence phase. In the ecotype 1 and 2 phases, biodiversity is merely the neutral diversity of the single ecotype that survives in habitat 2, and the system cannot support stable adaptive diversity. To demonstrate the robustness of results to the choice of biodiversity measure, we also plot Shannon entropy (Fig.~2c dotted red curves), which shows a close qualitative agreement with Renyi entropy. The conditions for phase transitions and critical migration rates are summarized in Table 1.

\begin{table}[b]
\caption{\label{table1}Conditions for phase transitions and critical migration rates, at which an ecotype goes extinct or emerges. (*) At $k=\lambda_1$, $\tau\to\infty$ always. However, in case (1) this does not indicate a genuine extinction, because $\vec\psi_1$ and $\vec\psi_2$ do not represent genetically distinct ecotypes.}
\begin{ruledtabular}
\begin{tabular}{lcr}
Extinction & Criteria & Critical $k$ \\
\colrule
Ecotype 1 (*) & Always occurs& $\lambda_1$  \\
Ecotype 2 & $\vec\psi_1\cdot\vec\psi_2=0$ \& $\begin{cases}\lambda_1 K_1\geq\eta \lambda_2 K_2\\ \lambda_1 K_1<\eta \lambda_2 K_2\end{cases}$ & $\begin{aligned}k_+\textrm{ and }k_- \\ \textrm{None} \end{aligned}$
\end{tabular}
\end{ruledtabular}
\end{table}

As Fig. 2 shows, the relationship between biodiversity and migration rate is largely determined by whether the critical migration rates $k_+$ and $k_-$ exist (i.e. whether $R_\textrm{max}>R_c$) and, if they do, by their values. One can tune $R_\textrm{max}$ and $R_c$ by adjusting the resource constraints. As a result, studying the phase transitions can provide useful insight into biodiversity and its sensitivity to migration rate.

\textit{Mechanism of Critical Slowing Down} --- The three phases discussed above are separated by transitions where an ecotype goes extinct or emerges as $k$ increases. At each transition, the dynamics exhibit critical slowing down (CSD, like in Fig.~1d, case 2). Such slowing relaxation is apparent from simulated dynamics. In the SM \cite{SM} we show simulations for $k$ below, at, and above each phase transition. One can see that, after early transients that depend on initial condition, populations gradually relax to steady state. 
At the phase transitions, CSD slows not only the extinction of a vanishing ecotype ($k=k_-$, Movie 2A) but also the growth of a re-emerging ecotype ($k\geq k_+$, Movies 8 and 9). On the other hand, the emergence of an ecotype via mutations from the other ecotype, through fitness valley crossing, can also be very slow, regardless of whether $k$ is near a critical value. This slow mutational process is distinct from the critical slowing down that we discuss here and is not our focus.
In what follows, we elucidate the mechanistic origin of CSD in our context, and discuss its implications for preserving biodiversity and detecting impending extinction.

Mathematically, CSD refers to a situation where an eigenvalue of the system's stability matrix (the Jacobian of the system of ODEs) vanishes as the control parameter reaches a critical value, resulting in slow (sub-exponential) relaxation to steady state. In our model, CSD occurs at each critical migration rate, i.e. $k=k_-$, $k_+$ or $\lambda_1$, if the proper criteria are satisfied (Table 1). Fig.~3a shows the migration rate dependence of the system's relaxation time $\tau$ (thick blue curve) defined as $\tau\equiv-1/w_1$, where $w_1$ is the largest (least negative) eigenvalue of the stability matrix; as $w_1$ approaches zero at each transition, $\tau$ peaks sharply (Fig.~3b). 

\begin{figure}[b]
\includegraphics[width=7cm]{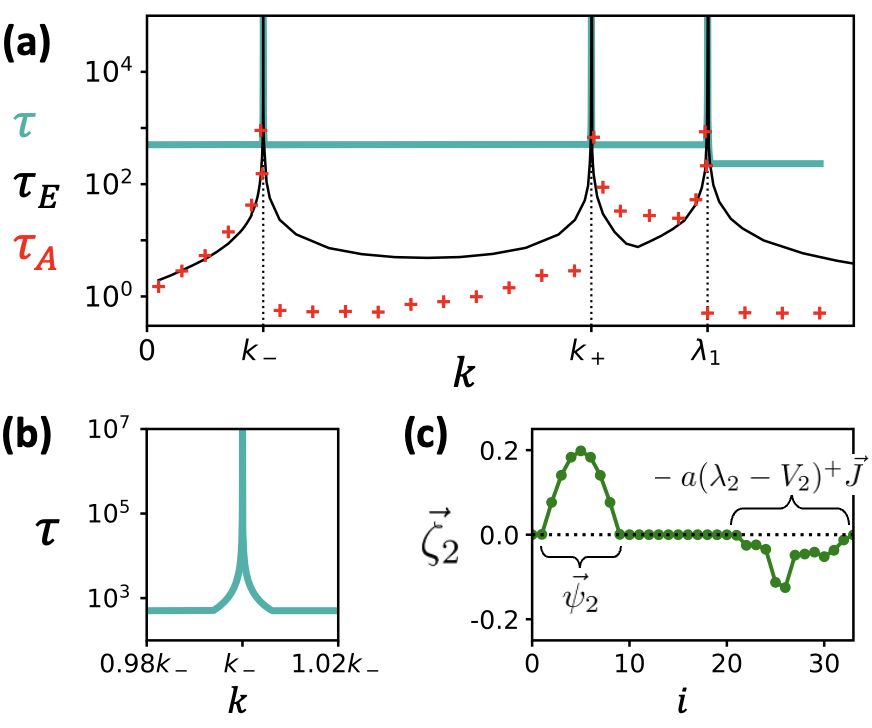}
\caption{\label{fig:1} 
\textbf{Resource competition between ecotypes causes critical slowing down.} \textbf{(a)} Relaxation time $\tau=-1/w_1$ (thick blue), ecological relaxation time $\tau_E$ (thin black), and time constant of autocorrelation decay $\tau_A$ (red marks). Vertical dashed lines indicate critical migration rates. Same setting as Fig.~1, case (2). \textbf{(b)} Close-up of the peak in $\tau$ at $k=k_-$. Similar peaks occur at $k=k_+$ and $k=\lambda_1$. \textbf{(c)} The zero mode in habitat 2, $\vec\zeta_2$. The two terms $\vec\psi_2$ and $-a(\lambda_2-\hat{V}_2)^+\vec J$ (indicated in the figure) comprise genotypes over ecotype 2 and ecotype 1, respectively. Due to the opposite sign, they describe a re-allocation of population from ecotype 2 to ecotype 1, as a result of resource competition.}
\end{figure}

Interestingly, the transition at $R=R_c$ ($k=k_-$ or $k_+$) occurs only in case (2), when the two ecotypes have zero overlap in genotype. This suggests that ecotypes compete as distinct ecological units, and that this competition might be the origin of CSD. To explore this possibility, we examine the slow relaxation dynamics of genotype abundances at $R=R_c$. At this critical flux, the stability matrix has a zero mode $\bm{\zeta}\equiv[\vec\zeta_1\:\, \vec\zeta_2]^T\in\mathbb{R}^{2M}$ (the superscript $^T$ denotes transpose) associated with the vanishing eigenvalue $w_1$. Deviations from the steady state along the zero mode decay sub-exponentially ($\sim 1/t$, as we will show), while those along any other eigenvector decay exponentially fast. 
Specifically, consider a deviation from the steady state along the zero mode: $[\vec n_1(t) \:\, \vec n_2(t)]^T=[\vec n^*_1 \:\, \vec n^*_2]^T+c(t)[\vec\zeta_1 \:\, \vec\zeta_2]^T$. As the coefficient $c(t)$ decays slowly to zero, $\vec\zeta_1$ and $\vec\zeta_2$ describe how the distribution of genotypes in habitats 1 and 2 change, respectively. We find that \cite{SM} $\vec\zeta_1=\vec{0}$ and
\begin{equation}
    \vec\zeta_2 = \vec\psi_2 - a(\lambda_2-\hat{V}_2)^+\vec J
\end{equation}
where $a$ is a constant; Fig.~3c shows $\vec\zeta_2$. Importantly, the two terms $\vec\psi_2$ and $a(\lambda_2-\hat{V}_2)^+\vec J$ represent clusters comprising the same genotypes as ecotype 2 and ecotype 1, but with opposite sign (Fig.~3c). Because of this, a decrease in the coefficient $c(t)$ causes the population of ecotype 1 to increase and the population of ecotype 2 to decrease. In other words, the zero mode $\bm{\zeta}$ describes a re-allocation of population from ecotype 2 to ecotype 1 within habitat 2.

The observation that the zero mode re-allocates population between the ecotypes is key to understanding the sub-exponential decay of $c(t)$. In particular, it suggests that the dynamics of $c(t)$ reflect ecological competition between ecotypes. Indeed, at $R=R_c$, the dynamics governed by Eqs.~\ref{eq:EOM1} and \ref{eq:EOM2} satisfy
\begin{equation}\label{dcdt}
    \frac{d}{dt}c(t)=-\rho_2\zeta_2^\textrm{tot} c(t)^2
\end{equation}
where $\zeta_2^\textrm{tot}$ is the sum of components of $\vec\zeta_2$. The significance of this result is that, under a critical flux, reproduction, mutation, and migration exactly balance out, leaving only the resource limitation driving the dynamics. In other words, resource competition, rather than evolution, governs the slow relaxation. This raises the question: does evolution play any role in the phase transitions? In fact, fitness landscapes and mutation rate do affect the characteristics of the ecotypes ($\vec\psi_l$, $\lambda_l$) as well as the value of $R_c$ (see Eq.~SM7). However, once these characteristics are set, CSD can be understood as an ecological process.

Solving Eq.~\ref{dcdt} yields $c(t)=1/(\rho_2\zeta_2^\textrm{tot} t+c_0^{-1})$ where $c_0$ is the initial magnitude of deviation, thus confirming a slow $1/t$ relaxation. A similar analysis can be done for the transition at $k=\lambda_1$, where the zero mode also reallocates population between the ecotypes and obeys $c(t)=1/(\rho_1\zeta_1^\textrm{tot} t+c_0^{-1})$~\cite{SM}. In fact, both transitions result from transcritical bifurcation in the steady-state populations, like those in a simple logistic model (see SM \cite{SM} for details). 
Note that in our evolutionary model, a balance of mutation, selection and migration only occurs at the critical flux, and this balance reduces the dynamics to being solely driven by a logistic term that describes ecological interaction between evolved ecotypes. 
Simulations in the SM~\cite{SM} compare the slow dynamics at transitions to the exponentially fast relaxation at other migration rates.

The existence of a zero mode at $k=k_-$, $k_+$ and $\lambda_1$ implies diverging $\tau$ at each critical $k$. However, away from the critical rates (flat portion of the thick blue curve in Fig.~3a), relaxation modes describe a different process insensitive to $k$: within-ecotype evolution~\cite{SM}. To study ecotype interaction away from the transitions, we consider the relaxation of each ecotype's total population in habitat 2, which we denote by $E_1\equiv\sum_{i\in G_1} n_{2,i}$ and $E_2\equiv\sum_{i\in G_2} n_{2,i}$, where $G_1$ and $G_2$ are the set of genotypes composing ecotypes 1 and 2, respectively. Starting from the native populations $\vec n_1=K_1\vec\psi_1$ and $\vec n_2=K_2\vec\psi_2$ (i.e. steady state populations with $k=0$), we simulate Eqs.~1 and 2 and compute the time constant by which $E_1$ and $E_2$ relax to their steady state values. We call this time constant the \textit{ecological relaxation time}, $\tau_E$ (Fig.~3a, thin black curve). As expected, $\tau$ and $\tau_E$ agree near the transitions but disagree far from the transitions where the relaxation modes describe within-ecotype evolution. The broader peaks in $\tau_E$ make them potentially easier to detect in experiment. 

From an ecological perspective, $\tau_E$, rather than $\tau$, might be of greater relevance, as it describes the dynamics of ecotype interaction while ignoring any within-ecotype evolution (i.e. changes in genotypic makeup). Furthermore, the behavior of $\tau_E$ suggests an intriguing ecological consequence of critical slowing down. Suppose two ecotypes coexist stably ($0<k<k_-$ or $k_+<k<\lambda_1$). As $k$ is varied to above $k_-$ or below $k_+$, ecotype 2 is no longer stable. The ecological relaxation time $\tau_E$ quantifies the timescale over which ecotype 2 approaches extinction. As a consequence, peaks in $\tau_E$ may act as a ``barrier" to extinction during transient variations in the migration rate; for instance, peaks in $\tau_E$ at $k_-$ and $k_+$ (Fig.~3b) could protect ecotype 2 from extinction.

To explore the effect of CSD in a more realistic setting, we introduce demographic noise by adding $\alpha\sqrt{\vec n_1}\xi_1(t)$ to Eq.~1 and $\alpha\sqrt{\vec n_2}\xi_2(t)$ to Eq.~2, where the square root is element-wise. $\xi_1(t)$ and $\xi_2(t)$ are noise terms that obey $\langle\xi_l(t)\rangle=0$ and $\langle\xi_l(t)\xi_l(t')\rangle=\delta(t-t')$ for $l=1,2$; the parameter $\alpha$ controls the strength of noise. Demographic noise causes the populations to fluctuate around their long-time averages. The statistics of these fluctuations encode the relaxation times of the system. To be more precise, let $\delta E_l$ be the deviation of ecotype $l$'s total population $E_l$ from its mean value. If $\delta E_l(t)$ were to obey a linear Langevin equation, then the time constant $\tau_A$ for the decay of the autocorrelation $\langle \delta E_l(0)\delta E_l(t) \rangle$ would equal the time constant $\tau_E$ for the relaxation of $E_l$ to its steady state value. In our model, this picture is complicated by the nonlinearity of the dynamics and the fact that multiple eigenmodes may contribute to relaxation. Nevertheless, $\tau_A$ provides a good estimate of $\tau_E$ in the coexistence phase (Fig.~3a, red dots). The close agreement suggests that measuring autocorrelation might be a viable approach for detecting impending extinction in experimental systems. In the single-ecotype phases, fluctuations due to demographic noise do not generate ecological competition because one ecotype is absent. Thus, $\tau_A$ is unrelated to $\tau_E$ in these phases (see SM \cite{SM} for further discussion). Although the evolutionary dynamics described by Eqs.~\ref{eq:EOM1} and \ref{eq:EOM2} are generally quite complex, the simple mechanism that causes critical slowing down permits a simple indicator $\tau_A$ that can be measured in noisy systems to reveal the phase transitions.

\textit{Discussion} --- In this work, we quantify the relationship between biodiversity and migration rate using a simple evolutionary model, aimed at elucidating to what extent spatial heterogeneity alone can maintain fine-scale diversity of closely-related species or strains. We find this relationship to be non-trivial even for one-way migration between two habitats. While diversity within an ecotype is independent of migration rate, diversity of the whole population is low for both low and high migration rates, with one or two maxima at intermediate rates. Ecotypes go extinct or emerge at critical migration rates, where critical slowing down results from resource competition between ecotypes. Such diverging relaxation times may act to protect biodiversity from transient variations in migration rate.  


Our results make a series of predictions testable by experiment (for example, using bioreactors/chemostats \cite{hansen2020antibiotics} or synthetic microenvironments \cite{zhang2011acceleration} as the connected habitats), even when the fitness landscapes are unknown. The only requirement is the ability to tune the one-way migration rate over a wide range and to measure the genotype makeup via sequencing. First, by probing if the native ecotypes of the two habitats share common genotypes, one can determine whether or not to expect phase transitions (signified by diverging relaxation times). Further, if transitions were expected, one can tune the number and location of transitions by varying resource supply (e.g. nutrient concentration or antigen dosage) to the provider habitat (tuning $R_\textrm{max}$) and/or the receiver habitat (adjusting $R_c$) so as to alter the relation between $R_\textrm{max}$ and $R_c$. Lastly, if $R_\textrm{max}>R_c$ is satisfied, one can linearly ramp up the migration rate (e.g. via controlling the flow speed) and seek non-monotonic changes in biodiversity calculated from genotype frequencies obtained by sequencing. The re-entrant transition can be detected from recurrence of the native ecotype in the receiver habitat at high migration rates, at which the decay of autocorrelations becomes sub-exponential.

While we have considered the simplest scenarios, the scope of the biodiversity-migration relationship of an evolving meta-population can be broadened considerably through two extensions. First, two-way migration would allow feedback between habitats and might result in ecotypes distinct from those native to any individual habitat. It will be interesting to seek signatures of feedback from features of phase transitions. Second, generalizing our analysis to a network of habitats/niches or a spatial continuum would allow a closer comparison with specific natural or laboratory systems, revealing ways in which spatial architecture of migration flux and selection profile might impact global biodiversity.  
Moreover, critical slowing down identified at particular migration rates might have important implications for the catastrophic onset of drug resistance in metastatic cancers, i.e. the takeover of adapted mutants in secondary tumors. Meanwhile, it might suggest potential means to sense and avoid tipping points by tuning the migration rate away from the critical values, via altering intrinsic mobility of cells (e.g. motor activity or metabolism) or tuning environmental properties (e.g. matrix stiffness or fiber alignment).

We thank Nandita Garud and Richard Wolff for illuminating discussions. CB is grateful for support from the NSF Graduate Research Fellowship (NSF Grant No. DGE-2034835). SW is supported by an NSF CAREER Award (Grant No. PHY-2146581). 

\bibliography{citations}

\begin{thebibliography}{36}%
\makeatletter
\providecommand \@ifxundefined [1]{%
 \@ifx{#1\undefined}
}%
\providecommand \@ifnum [1]{%
 \ifnum #1\expandafter \@firstoftwo
 \else \expandafter \@secondoftwo
 \fi
}%
\providecommand \@ifx [1]{%
 \ifx #1\expandafter \@firstoftwo
 \else \expandafter \@secondoftwo
 \fi
}%
\providecommand \natexlab [1]{#1}%
\providecommand \enquote  [1]{``#1''}%
\providecommand \bibnamefont  [1]{#1}%
\providecommand \bibfnamefont [1]{#1}%
\providecommand \citenamefont [1]{#1}%
\providecommand \href@noop [0]{\@secondoftwo}%
\providecommand \href [0]{\begingroup \@sanitize@url \@href}%
\providecommand \@href[1]{\@@startlink{#1}\@@href}%
\providecommand \@@href[1]{\endgroup#1\@@endlink}%
\providecommand \@sanitize@url [0]{\catcode `\\12\catcode `\$12\catcode
  `\&12\catcode `\#12\catcode `\^12\catcode `\_12\catcode `\%12\relax}%
\providecommand \@@startlink[1]{}%
\providecommand \@@endlink[0]{}%
\providecommand \url  [0]{\begingroup\@sanitize@url \@url }%
\providecommand \@url [1]{\endgroup\@href {#1}{\urlprefix }}%
\providecommand \urlprefix  [0]{URL }%
\providecommand \Eprint [0]{\href }%
\providecommand \doibase [0]{http://dx.doi.org/}%
\providecommand \selectlanguage [0]{\@gobble}%
\providecommand \bibinfo  [0]{\@secondoftwo}%
\providecommand \bibfield  [0]{\@secondoftwo}%
\providecommand \translation [1]{[#1]}%
\providecommand \BibitemOpen [0]{}%
\providecommand \bibitemStop [0]{}%
\providecommand \bibitemNoStop [0]{.\EOS\space}%
\providecommand \EOS [0]{\spacefactor3000\relax}%
\providecommand \BibitemShut  [1]{\csname bibitem#1\endcsname}%
\let\auto@bib@innerbib\@empty
\bibitem [{\citenamefont {Garud}\ \emph {et~al.}(2019)\citenamefont {Garud},
  \citenamefont {Good}, \citenamefont {Hallatschek},\ and\ \citenamefont
  {Pollard}}]{garud2019evolutionary}%
  \BibitemOpen
  \bibfield  {author} {\bibinfo {author} {\bibfnamefont {N.~R.}\ \bibnamefont
  {Garud}}, \bibinfo {author} {\bibfnamefont {B.~H.}\ \bibnamefont {Good}},
  \bibinfo {author} {\bibfnamefont {O.}~\bibnamefont {Hallatschek}}, \ and\
  \bibinfo {author} {\bibfnamefont {K.~S.}\ \bibnamefont {Pollard}},\
  }\href@noop {} {\bibfield  {journal} {\bibinfo  {journal} {PLoS biology}\
  }\textbf {\bibinfo {volume} {17}},\ \bibinfo {pages} {e3000102} (\bibinfo
  {year} {2019})}\BibitemShut {NoStop}%
\bibitem [{\citenamefont {Lieberman}\ \emph {et~al.}(2016)\citenamefont
  {Lieberman}, \citenamefont {Wilson}, \citenamefont {Misra}, \citenamefont
  {Xiong}, \citenamefont {Moodley}, \citenamefont {Cohen},\ and\ \citenamefont
  {Kishony}}]{lieberman2016}%
  \BibitemOpen
  \bibfield  {author} {\bibinfo {author} {\bibfnamefont {T.~D.}\ \bibnamefont
  {Lieberman}}, \bibinfo {author} {\bibfnamefont {D.}~\bibnamefont {Wilson}},
  \bibinfo {author} {\bibfnamefont {R.}~\bibnamefont {Misra}}, \bibinfo
  {author} {\bibfnamefont {L.~L.}\ \bibnamefont {Xiong}}, \bibinfo {author}
  {\bibfnamefont {P.}~\bibnamefont {Moodley}}, \bibinfo {author} {\bibfnamefont
  {T.}~\bibnamefont {Cohen}}, \ and\ \bibinfo {author} {\bibfnamefont
  {R.}~\bibnamefont {Kishony}},\ }\href@noop {} {\bibfield  {journal} {\bibinfo
   {journal} {Nature medicine}\ }\textbf {\bibinfo {volume} {22}},\ \bibinfo
  {pages} {1470} (\bibinfo {year} {2016})}\BibitemShut {NoStop}%
\bibitem [{\citenamefont {Oh}\ \emph {et~al.}(2016)\citenamefont {Oh},
  \citenamefont {Byrd}, \citenamefont {Park}, \citenamefont {Kong},
  \citenamefont {Segre}, \citenamefont {Program} \emph
  {et~al.}}]{oh2016temporal}%
  \BibitemOpen
  \bibfield  {author} {\bibinfo {author} {\bibfnamefont {J.}~\bibnamefont
  {Oh}}, \bibinfo {author} {\bibfnamefont {A.~L.}\ \bibnamefont {Byrd}},
  \bibinfo {author} {\bibfnamefont {M.}~\bibnamefont {Park}}, \bibinfo {author}
  {\bibfnamefont {H.~H.}\ \bibnamefont {Kong}}, \bibinfo {author}
  {\bibfnamefont {J.~A.}\ \bibnamefont {Segre}}, \bibinfo {author}
  {\bibfnamefont {N.~C.~S.}\ \bibnamefont {Program}},  \emph {et~al.},\
  }\href@noop {} {\bibfield  {journal} {\bibinfo  {journal} {Cell}\ }\textbf
  {\bibinfo {volume} {165}},\ \bibinfo {pages} {854} (\bibinfo {year}
  {2016})}\BibitemShut {NoStop}%
\bibitem [{\citenamefont {Tikhonov}\ \emph {et~al.}(2015)\citenamefont
  {Tikhonov}, \citenamefont {Leach},\ and\ \citenamefont
  {Wingreen}}]{tikhonov2015interpreting}%
  \BibitemOpen
  \bibfield  {author} {\bibinfo {author} {\bibfnamefont {M.}~\bibnamefont
  {Tikhonov}}, \bibinfo {author} {\bibfnamefont {R.~W.}\ \bibnamefont {Leach}},
  \ and\ \bibinfo {author} {\bibfnamefont {N.~S.}\ \bibnamefont {Wingreen}},\
  }\href@noop {} {\bibfield  {journal} {\bibinfo  {journal} {The ISME journal}\
  }\textbf {\bibinfo {volume} {9}},\ \bibinfo {pages} {68} (\bibinfo {year}
  {2015})}\BibitemShut {NoStop}%
\bibitem [{\citenamefont {Acinas}\ \emph {et~al.}(2004)\citenamefont {Acinas},
  \citenamefont {Klepac-Ceraj}, \citenamefont {Hunt}, \citenamefont {Pharino},
  \citenamefont {Ceraj}, \citenamefont {Distel},\ and\ \citenamefont
  {Polz}}]{acinas2004fine}%
  \BibitemOpen
  \bibfield  {author} {\bibinfo {author} {\bibfnamefont {S.~G.}\ \bibnamefont
  {Acinas}}, \bibinfo {author} {\bibfnamefont {V.}~\bibnamefont
  {Klepac-Ceraj}}, \bibinfo {author} {\bibfnamefont {D.~E.}\ \bibnamefont
  {Hunt}}, \bibinfo {author} {\bibfnamefont {C.}~\bibnamefont {Pharino}},
  \bibinfo {author} {\bibfnamefont {I.}~\bibnamefont {Ceraj}}, \bibinfo
  {author} {\bibfnamefont {D.~L.}\ \bibnamefont {Distel}}, \ and\ \bibinfo
  {author} {\bibfnamefont {M.~F.}\ \bibnamefont {Polz}},\ }\href@noop {}
  {\bibfield  {journal} {\bibinfo  {journal} {Nature}\ }\textbf {\bibinfo
  {volume} {430}},\ \bibinfo {pages} {551} (\bibinfo {year}
  {2004})}\BibitemShut {NoStop}%
\bibitem [{\citenamefont {Rosen}\ \emph {et~al.}(2018)\citenamefont {Rosen},
  \citenamefont {Davison}, \citenamefont {Fisher},\ and\ \citenamefont
  {Bhaya}}]{rosen2018probing}%
  \BibitemOpen
  \bibfield  {author} {\bibinfo {author} {\bibfnamefont {M.~J.}\ \bibnamefont
  {Rosen}}, \bibinfo {author} {\bibfnamefont {M.}~\bibnamefont {Davison}},
  \bibinfo {author} {\bibfnamefont {D.~S.}\ \bibnamefont {Fisher}}, \ and\
  \bibinfo {author} {\bibfnamefont {D.}~\bibnamefont {Bhaya}},\ }\href@noop {}
  {\bibfield  {journal} {\bibinfo  {journal} {PloS one}\ }\textbf {\bibinfo
  {volume} {13}},\ \bibinfo {pages} {e0205396} (\bibinfo {year}
  {2018})}\BibitemShut {NoStop}%
\bibitem [{\citenamefont {Rosen}\ \emph {et~al.}(2015)\citenamefont {Rosen},
  \citenamefont {Davison}, \citenamefont {Bhaya},\ and\ \citenamefont
  {Fisher}}]{rosen2015fine}%
  \BibitemOpen
  \bibfield  {author} {\bibinfo {author} {\bibfnamefont {M.~J.}\ \bibnamefont
  {Rosen}}, \bibinfo {author} {\bibfnamefont {M.}~\bibnamefont {Davison}},
  \bibinfo {author} {\bibfnamefont {D.}~\bibnamefont {Bhaya}}, \ and\ \bibinfo
  {author} {\bibfnamefont {D.~S.}\ \bibnamefont {Fisher}},\ }\href@noop {}
  {\bibfield  {journal} {\bibinfo  {journal} {Science}\ }\textbf {\bibinfo
  {volume} {348}},\ \bibinfo {pages} {1019} (\bibinfo {year}
  {2015})}\BibitemShut {NoStop}%
\bibitem [{\citenamefont {Conwill}\ \emph {et~al.}(2022)\citenamefont
  {Conwill}, \citenamefont {Kuan}, \citenamefont {Damerla}, \citenamefont
  {Poret}, \citenamefont {Baker}, \citenamefont {Tripp}, \citenamefont {Alm},\
  and\ \citenamefont {Lieberman}}]{conwill2022}%
  \BibitemOpen
  \bibfield  {author} {\bibinfo {author} {\bibfnamefont {A.}~\bibnamefont
  {Conwill}}, \bibinfo {author} {\bibfnamefont {A.~C.}\ \bibnamefont {Kuan}},
  \bibinfo {author} {\bibfnamefont {R.}~\bibnamefont {Damerla}}, \bibinfo
  {author} {\bibfnamefont {A.~J.}\ \bibnamefont {Poret}}, \bibinfo {author}
  {\bibfnamefont {J.~S.}\ \bibnamefont {Baker}}, \bibinfo {author}
  {\bibfnamefont {A.~D.}\ \bibnamefont {Tripp}}, \bibinfo {author}
  {\bibfnamefont {E.~J.}\ \bibnamefont {Alm}}, \ and\ \bibinfo {author}
  {\bibfnamefont {T.~D.}\ \bibnamefont {Lieberman}},\ }\href@noop {} {\bibfield
   {journal} {\bibinfo  {journal} {Cell Host \& Microbe}\ }\textbf {\bibinfo
  {volume} {30}},\ \bibinfo {pages} {171} (\bibinfo {year} {2022})}\BibitemShut
  {NoStop}%
\bibitem [{\citenamefont {Hutchinson}(1961)}]{hutchinson1961paradox}%
  \BibitemOpen
  \bibfield  {author} {\bibinfo {author} {\bibfnamefont {G.~E.}\ \bibnamefont
  {Hutchinson}},\ }\href@noop {} {\bibfield  {journal} {\bibinfo  {journal}
  {The American Naturalist}\ }\textbf {\bibinfo {volume} {95}},\ \bibinfo
  {pages} {137} (\bibinfo {year} {1961})}\BibitemShut {NoStop}%
\bibitem [{\citenamefont {Chesson}(2000)}]{chesson2000mechanisms}%
  \BibitemOpen
  \bibfield  {author} {\bibinfo {author} {\bibfnamefont {P.}~\bibnamefont
  {Chesson}},\ }\href@noop {} {\bibfield  {journal} {\bibinfo  {journal}
  {Annual review of Ecology and Systematics}\ }\textbf {\bibinfo {volume}
  {31}},\ \bibinfo {pages} {343} (\bibinfo {year} {2000})}\BibitemShut
  {NoStop}%
\bibitem [{\citenamefont {Amarasekare}(2003)}]{amarasekare2003competitive}%
  \BibitemOpen
  \bibfield  {author} {\bibinfo {author} {\bibfnamefont {P.}~\bibnamefont
  {Amarasekare}},\ }\href@noop {} {\bibfield  {journal} {\bibinfo  {journal}
  {Ecology letters}\ }\textbf {\bibinfo {volume} {6}},\ \bibinfo {pages} {1109}
  (\bibinfo {year} {2003})}\BibitemShut {NoStop}%
\bibitem [{\citenamefont {Ellner}\ \emph {et~al.}(2019)\citenamefont {Ellner},
  \citenamefont {Snyder}, \citenamefont {Adler},\ and\ \citenamefont
  {Hooker}}]{ellner2019expanded}%
  \BibitemOpen
  \bibfield  {author} {\bibinfo {author} {\bibfnamefont {S.~P.}\ \bibnamefont
  {Ellner}}, \bibinfo {author} {\bibfnamefont {R.~E.}\ \bibnamefont {Snyder}},
  \bibinfo {author} {\bibfnamefont {P.~B.}\ \bibnamefont {Adler}}, \ and\
  \bibinfo {author} {\bibfnamefont {G.}~\bibnamefont {Hooker}},\ }\href@noop {}
  {\bibfield  {journal} {\bibinfo  {journal} {Ecology letters}\ }\textbf
  {\bibinfo {volume} {22}},\ \bibinfo {pages} {3} (\bibinfo {year}
  {2019})}\BibitemShut {NoStop}%
\bibitem [{\citenamefont {Pearce}\ \emph {et~al.}(2020)\citenamefont {Pearce},
  \citenamefont {Agarwala},\ and\ \citenamefont
  {Fisher}}]{pearce2020stabilization}%
  \BibitemOpen
  \bibfield  {author} {\bibinfo {author} {\bibfnamefont {M.~T.}\ \bibnamefont
  {Pearce}}, \bibinfo {author} {\bibfnamefont {A.}~\bibnamefont {Agarwala}}, \
  and\ \bibinfo {author} {\bibfnamefont {D.~S.}\ \bibnamefont {Fisher}},\
  }\href@noop {} {\bibfield  {journal} {\bibinfo  {journal} {Proceedings of the
  National Academy of Sciences}\ }\textbf {\bibinfo {volume} {117}},\ \bibinfo
  {pages} {14572} (\bibinfo {year} {2020})}\BibitemShut {NoStop}%
\bibitem [{\citenamefont {Roy}\ \emph {et~al.}(2020)\citenamefont {Roy},
  \citenamefont {Barbier}, \citenamefont {Biroli},\ and\ \citenamefont
  {Bunin}}]{roy2020complex}%
  \BibitemOpen
  \bibfield  {author} {\bibinfo {author} {\bibfnamefont {F.}~\bibnamefont
  {Roy}}, \bibinfo {author} {\bibfnamefont {M.}~\bibnamefont {Barbier}},
  \bibinfo {author} {\bibfnamefont {G.}~\bibnamefont {Biroli}}, \ and\ \bibinfo
  {author} {\bibfnamefont {G.}~\bibnamefont {Bunin}},\ }\href@noop {}
  {\bibfield  {journal} {\bibinfo  {journal} {PLoS computational biology}\
  }\textbf {\bibinfo {volume} {16}},\ \bibinfo {pages} {e1007827} (\bibinfo
  {year} {2020})}\BibitemShut {NoStop}%
\bibitem [{\citenamefont {Hart}\ \emph {et~al.}(2017)\citenamefont {Hart},
  \citenamefont {Usinowicz},\ and\ \citenamefont {Levine}}]{hart2017spatial}%
  \BibitemOpen
  \bibfield  {author} {\bibinfo {author} {\bibfnamefont {S.~P.}\ \bibnamefont
  {Hart}}, \bibinfo {author} {\bibfnamefont {J.}~\bibnamefont {Usinowicz}}, \
  and\ \bibinfo {author} {\bibfnamefont {J.~M.}\ \bibnamefont {Levine}},\
  }\href@noop {} {\bibfield  {journal} {\bibinfo  {journal} {Nature Ecology \&
  Evolution}\ }\textbf {\bibinfo {volume} {1}},\ \bibinfo {pages} {1066}
  (\bibinfo {year} {2017})}\BibitemShut {NoStop}%
\bibitem [{\citenamefont {Good}\ and\ \citenamefont
  {Hallatschek}(2018)}]{good2018effective}%
  \BibitemOpen
  \bibfield  {author} {\bibinfo {author} {\bibfnamefont {B.~H.}\ \bibnamefont
  {Good}}\ and\ \bibinfo {author} {\bibfnamefont {O.}~\bibnamefont
  {Hallatschek}},\ }\href@noop {} {\bibfield  {journal} {\bibinfo  {journal}
  {Current opinion in microbiology}\ }\textbf {\bibinfo {volume} {45}},\
  \bibinfo {pages} {203} (\bibinfo {year} {2018})}\BibitemShut {NoStop}%
\bibitem [{\citenamefont {Dias}(1996)}]{dias1996sources}%
  \BibitemOpen
  \bibfield  {author} {\bibinfo {author} {\bibfnamefont {P.~C.}\ \bibnamefont
  {Dias}},\ }\href@noop {} {\bibfield  {journal} {\bibinfo  {journal} {Trends
  in Ecology \& Evolution}\ }\textbf {\bibinfo {volume} {11}},\ \bibinfo
  {pages} {326} (\bibinfo {year} {1996})}\BibitemShut {NoStop}%
\bibitem [{\citenamefont {Hermsen}\ and\ \citenamefont
  {Hwa}(2010)}]{hermsen2010sources}%
  \BibitemOpen
  \bibfield  {author} {\bibinfo {author} {\bibfnamefont {R.}~\bibnamefont
  {Hermsen}}\ and\ \bibinfo {author} {\bibfnamefont {T.}~\bibnamefont {Hwa}},\
  }\href@noop {} {\bibfield  {journal} {\bibinfo  {journal} {Physical review
  letters}\ }\textbf {\bibinfo {volume} {105}},\ \bibinfo {pages} {248104}
  (\bibinfo {year} {2010})}\BibitemShut {NoStop}%
\bibitem [{\citenamefont {Hermsen}\ \emph {et~al.}(2012)\citenamefont
  {Hermsen}, \citenamefont {Deris},\ and\ \citenamefont
  {Hwa}}]{hermsen2012rapidity}%
  \BibitemOpen
  \bibfield  {author} {\bibinfo {author} {\bibfnamefont {R.}~\bibnamefont
  {Hermsen}}, \bibinfo {author} {\bibfnamefont {J.~B.}\ \bibnamefont {Deris}},
  \ and\ \bibinfo {author} {\bibfnamefont {T.}~\bibnamefont {Hwa}},\
  }\href@noop {} {\bibfield  {journal} {\bibinfo  {journal} {Proceedings of the
  National Academy of Sciences}\ }\textbf {\bibinfo {volume} {109}},\ \bibinfo
  {pages} {10775} (\bibinfo {year} {2012})}\BibitemShut {NoStop}%
\bibitem [{\citenamefont {Zhao}\ \emph {et~al.}(2019)\citenamefont {Zhao},
  \citenamefont {Lieberman}, \citenamefont {Poyet}, \citenamefont {Kauffman},
  \citenamefont {Gibbons}, \citenamefont {Groussin}, \citenamefont {Xavier},\
  and\ \citenamefont {Alm}}]{zhao2019}%
  \BibitemOpen
  \bibfield  {author} {\bibinfo {author} {\bibfnamefont {S.}~\bibnamefont
  {Zhao}}, \bibinfo {author} {\bibfnamefont {T.~D.}\ \bibnamefont {Lieberman}},
  \bibinfo {author} {\bibfnamefont {M.}~\bibnamefont {Poyet}}, \bibinfo
  {author} {\bibfnamefont {K.~M.}\ \bibnamefont {Kauffman}}, \bibinfo {author}
  {\bibfnamefont {S.~M.}\ \bibnamefont {Gibbons}}, \bibinfo {author}
  {\bibfnamefont {M.}~\bibnamefont {Groussin}}, \bibinfo {author}
  {\bibfnamefont {R.~J.}\ \bibnamefont {Xavier}}, \ and\ \bibinfo {author}
  {\bibfnamefont {E.~J.}\ \bibnamefont {Alm}},\ }\href@noop {} {\bibfield
  {journal} {\bibinfo  {journal} {Cell host \& microbe}\ }\textbf {\bibinfo
  {volume} {25}},\ \bibinfo {pages} {656} (\bibinfo {year} {2019})}\BibitemShut
  {NoStop}%
\bibitem [{\citenamefont {Wolff}\ \emph {et~al.}(2021)\citenamefont {Wolff},
  \citenamefont {Shoemaker},\ and\ \citenamefont
  {Garud}}]{wolff2021ecological}%
  \BibitemOpen
  \bibfield  {author} {\bibinfo {author} {\bibfnamefont {R.}~\bibnamefont
  {Wolff}}, \bibinfo {author} {\bibfnamefont {W.~R.}\ \bibnamefont
  {Shoemaker}}, \ and\ \bibinfo {author} {\bibfnamefont {N.~R.}\ \bibnamefont
  {Garud}},\ }\href@noop {} {\bibfield  {journal} {\bibinfo  {journal}
  {bioRxiv}\ } (\bibinfo {year} {2021})}\BibitemShut {NoStop}%
\bibitem [{\citenamefont {Shai}\ \emph {et~al.}(2020)\citenamefont {Shai},
  \citenamefont {Alcal{\'a}-Corona}, \citenamefont {Wang}, \citenamefont {Kim},
  \citenamefont {Maslov}, \citenamefont {Whitaker},\ and\ \citenamefont
  {Pascual}}]{shai2020}%
  \BibitemOpen
  \bibfield  {author} {\bibinfo {author} {\bibfnamefont {P.}~\bibnamefont
  {Shai}}, \bibinfo {author} {\bibfnamefont {S.~A.}\ \bibnamefont
  {Alcal{\'a}-Corona}}, \bibinfo {author} {\bibfnamefont {T.}~\bibnamefont
  {Wang}}, \bibinfo {author} {\bibfnamefont {T.}~\bibnamefont {Kim}}, \bibinfo
  {author} {\bibfnamefont {S.}~\bibnamefont {Maslov}}, \bibinfo {author}
  {\bibfnamefont {R.}~\bibnamefont {Whitaker}}, \ and\ \bibinfo {author}
  {\bibfnamefont {M.}~\bibnamefont {Pascual}},\ }\href@noop {} {\bibfield
  {journal} {\bibinfo  {journal} {Nature Ecology \& Evolution}\ }\textbf
  {\bibinfo {volume} {4}},\ \bibinfo {pages} {1650} (\bibinfo {year}
  {2020})}\BibitemShut {NoStop}%
\bibitem [{\citenamefont {Sheng}\ and\ \citenamefont {Wang}(2021)}]{sheng2021}%
  \BibitemOpen
  \bibfield  {author} {\bibinfo {author} {\bibfnamefont {J.}~\bibnamefont
  {Sheng}}\ and\ \bibinfo {author} {\bibfnamefont {S.}~\bibnamefont {Wang}},\
  }\href@noop {} {\bibfield  {journal} {\bibinfo  {journal} {Iscience}\
  }\textbf {\bibinfo {volume} {24}},\ \bibinfo {pages} {102861} (\bibinfo
  {year} {2021})}\BibitemShut {NoStop}%
\bibitem [{\citenamefont {Arnoldini}\ \emph {et~al.}(2018)\citenamefont
  {Arnoldini}, \citenamefont {Cremer},\ and\ \citenamefont
  {Hwa}}]{arnoldini2018bacterial}%
  \BibitemOpen
  \bibfield  {author} {\bibinfo {author} {\bibfnamefont {M.}~\bibnamefont
  {Arnoldini}}, \bibinfo {author} {\bibfnamefont {J.}~\bibnamefont {Cremer}}, \
  and\ \bibinfo {author} {\bibfnamefont {T.}~\bibnamefont {Hwa}},\ }\href@noop
  {} {\bibfield  {journal} {\bibinfo  {journal} {Gut microbes}\ }\textbf
  {\bibinfo {volume} {9}},\ \bibinfo {pages} {559} (\bibinfo {year}
  {2018})}\BibitemShut {NoStop}%
\bibitem [{\citenamefont {Laichalk}\ \emph {et~al.}(2002)\citenamefont
  {Laichalk}, \citenamefont {Hochberg}, \citenamefont {Babcock}, \citenamefont
  {Freeman},\ and\ \citenamefont {Thorley-Lawson}}]{laichalk2002}%
  \BibitemOpen
  \bibfield  {author} {\bibinfo {author} {\bibfnamefont {L.~L.}\ \bibnamefont
  {Laichalk}}, \bibinfo {author} {\bibfnamefont {D.}~\bibnamefont {Hochberg}},
  \bibinfo {author} {\bibfnamefont {G.~J.}\ \bibnamefont {Babcock}}, \bibinfo
  {author} {\bibfnamefont {R.~B.}\ \bibnamefont {Freeman}}, \ and\ \bibinfo
  {author} {\bibfnamefont {D.~A.}\ \bibnamefont {Thorley-Lawson}},\ }\href@noop
  {} {\bibfield  {journal} {\bibinfo  {journal} {Immunity}\ }\textbf {\bibinfo
  {volume} {16}},\ \bibinfo {pages} {745} (\bibinfo {year} {2002})}\BibitemShut
  {NoStop}%
\bibitem [{\citenamefont {Bende}\ \emph {et~al.}(2007)\citenamefont {Bende},
  \citenamefont {Van~Maldegem}, \citenamefont {Triesscheijn}, \citenamefont
  {Wormhoudt}, \citenamefont {Guijt},\ and\ \citenamefont
  {Van~Noesel}}]{bende2007}%
  \BibitemOpen
  \bibfield  {author} {\bibinfo {author} {\bibfnamefont {R.~J.}\ \bibnamefont
  {Bende}}, \bibinfo {author} {\bibfnamefont {F.}~\bibnamefont {Van~Maldegem}},
  \bibinfo {author} {\bibfnamefont {M.}~\bibnamefont {Triesscheijn}}, \bibinfo
  {author} {\bibfnamefont {T.~A.}\ \bibnamefont {Wormhoudt}}, \bibinfo {author}
  {\bibfnamefont {R.}~\bibnamefont {Guijt}}, \ and\ \bibinfo {author}
  {\bibfnamefont {C.~J.}\ \bibnamefont {Van~Noesel}},\ }\href@noop {}
  {\bibfield  {journal} {\bibinfo  {journal} {The Journal of experimental
  medicine}\ }\textbf {\bibinfo {volume} {204}},\ \bibinfo {pages} {2655}
  (\bibinfo {year} {2007})}\BibitemShut {NoStop}%
\bibitem [{\citenamefont {Waclaw}\ \emph {et~al.}(2010)\citenamefont {Waclaw},
  \citenamefont {Allen},\ and\ \citenamefont {Evans}}]{waclaw2010dynamical}%
  \BibitemOpen
  \bibfield  {author} {\bibinfo {author} {\bibfnamefont {B.}~\bibnamefont
  {Waclaw}}, \bibinfo {author} {\bibfnamefont {R.~J.}\ \bibnamefont {Allen}}, \
  and\ \bibinfo {author} {\bibfnamefont {M.~R.}\ \bibnamefont {Evans}},\
  }\href@noop {} {\bibfield  {journal} {\bibinfo  {journal} {Physical review
  letters}\ }\textbf {\bibinfo {volume} {105}},\ \bibinfo {pages} {268101}
  (\bibinfo {year} {2010})}\BibitemShut {NoStop}%
\bibitem [{\citenamefont {Wissel}(1984)}]{wissel1984universal}%
  \BibitemOpen
  \bibfield  {author} {\bibinfo {author} {\bibfnamefont {C.}~\bibnamefont
  {Wissel}},\ }\href@noop {} {\bibfield  {journal} {\bibinfo  {journal}
  {Oecologia}\ }\textbf {\bibinfo {volume} {65}},\ \bibinfo {pages} {101}
  (\bibinfo {year} {1984})}\BibitemShut {NoStop}%
\bibitem [{\citenamefont {Drake}\ and\ \citenamefont
  {Griffen}(2010)}]{drake2010early}%
  \BibitemOpen
  \bibfield  {author} {\bibinfo {author} {\bibfnamefont {J.~M.}\ \bibnamefont
  {Drake}}\ and\ \bibinfo {author} {\bibfnamefont {B.~D.}\ \bibnamefont
  {Griffen}},\ }\href@noop {} {\bibfield  {journal} {\bibinfo  {journal}
  {Nature}\ }\textbf {\bibinfo {volume} {467}},\ \bibinfo {pages} {456}
  (\bibinfo {year} {2010})}\BibitemShut {NoStop}%
\bibitem [{\citenamefont {Dai}\ \emph {et~al.}(2012)\citenamefont {Dai},
  \citenamefont {Vorselen}, \citenamefont {Korolev},\ and\ \citenamefont
  {Gore}}]{dai2012generic}%
  \BibitemOpen
  \bibfield  {author} {\bibinfo {author} {\bibfnamefont {L.}~\bibnamefont
  {Dai}}, \bibinfo {author} {\bibfnamefont {D.}~\bibnamefont {Vorselen}},
  \bibinfo {author} {\bibfnamefont {K.~S.}\ \bibnamefont {Korolev}}, \ and\
  \bibinfo {author} {\bibfnamefont {J.}~\bibnamefont {Gore}},\ }\href@noop {}
  {\bibfield  {journal} {\bibinfo  {journal} {Science}\ }\textbf {\bibinfo
  {volume} {336}},\ \bibinfo {pages} {1175} (\bibinfo {year}
  {2012})}\BibitemShut {NoStop}%
\bibitem [{\citenamefont {Scheffer}\ \emph {et~al.}(2015)\citenamefont
  {Scheffer}, \citenamefont {Carpenter}, \citenamefont {Dakos},\ and\
  \citenamefont {van Nes}}]{scheffer2015generic}%
  \BibitemOpen
  \bibfield  {author} {\bibinfo {author} {\bibfnamefont {M.}~\bibnamefont
  {Scheffer}}, \bibinfo {author} {\bibfnamefont {S.~R.}\ \bibnamefont
  {Carpenter}}, \bibinfo {author} {\bibfnamefont {V.}~\bibnamefont {Dakos}}, \
  and\ \bibinfo {author} {\bibfnamefont {E.~H.}\ \bibnamefont {van Nes}},\
  }\href@noop {} {\bibfield  {journal} {\bibinfo  {journal} {Annual Review of
  Ecology, Evolution, and Systematics}\ }\textbf {\bibinfo {volume} {46}},\
  \bibinfo {pages} {145} (\bibinfo {year} {2015})}\BibitemShut {NoStop}%
\bibitem [{\citenamefont {Dakos}\ \emph {et~al.}(2019)\citenamefont {Dakos},
  \citenamefont {Matthews}, \citenamefont {Hendry}, \citenamefont {Levine},
  \citenamefont {Loeuille}, \citenamefont {Norberg}, \citenamefont {Nosil},
  \citenamefont {Scheffer},\ and\ \citenamefont
  {De~Meester}}]{dakos2019ecosystem}%
  \BibitemOpen
  \bibfield  {author} {\bibinfo {author} {\bibfnamefont {V.}~\bibnamefont
  {Dakos}}, \bibinfo {author} {\bibfnamefont {B.}~\bibnamefont {Matthews}},
  \bibinfo {author} {\bibfnamefont {A.~P.}\ \bibnamefont {Hendry}}, \bibinfo
  {author} {\bibfnamefont {J.}~\bibnamefont {Levine}}, \bibinfo {author}
  {\bibfnamefont {N.}~\bibnamefont {Loeuille}}, \bibinfo {author}
  {\bibfnamefont {J.}~\bibnamefont {Norberg}}, \bibinfo {author} {\bibfnamefont
  {P.}~\bibnamefont {Nosil}}, \bibinfo {author} {\bibfnamefont
  {M.}~\bibnamefont {Scheffer}}, \ and\ \bibinfo {author} {\bibfnamefont
  {L.}~\bibnamefont {De~Meester}},\ }\href@noop {} {\bibfield  {journal}
  {\bibinfo  {journal} {Nature ecology \& evolution}\ }\textbf {\bibinfo
  {volume} {3}},\ \bibinfo {pages} {355} (\bibinfo {year} {2019})}\BibitemShut
  {NoStop}%
\bibitem [{SM()}]{SM}%
  \BibitemOpen
  \href@noop {} {\bibinfo  {journal} {See Supplemental Material at [URL to be
  inserted]}\ }\BibitemShut {NoStop}%
\bibitem [{\citenamefont {Faith}\ \emph {et~al.}(2013)\citenamefont {Faith},
  \citenamefont {Guruge}, \citenamefont {Charbonneau}, \citenamefont
  {Subramanian}, \citenamefont {Seedorf}, \citenamefont {Goodman},
  \citenamefont {Clemente}, \citenamefont {Knight}, \citenamefont {Heath},
  \citenamefont {Leibel} \emph {et~al.}}]{faith2013long}%
  \BibitemOpen
\bibfield  {journal} {  }\bibfield  {author} {\bibinfo {author} {\bibfnamefont
  {J.~J.}\ \bibnamefont {Faith}}, \bibinfo {author} {\bibfnamefont {J.~L.}\
  \bibnamefont {Guruge}}, \bibinfo {author} {\bibfnamefont {M.}~\bibnamefont
  {Charbonneau}}, \bibinfo {author} {\bibfnamefont {S.}~\bibnamefont
  {Subramanian}}, \bibinfo {author} {\bibfnamefont {H.}~\bibnamefont
  {Seedorf}}, \bibinfo {author} {\bibfnamefont {A.~L.}\ \bibnamefont
  {Goodman}}, \bibinfo {author} {\bibfnamefont {J.~C.}\ \bibnamefont
  {Clemente}}, \bibinfo {author} {\bibfnamefont {R.}~\bibnamefont {Knight}},
  \bibinfo {author} {\bibfnamefont {A.~C.}\ \bibnamefont {Heath}}, \bibinfo
  {author} {\bibfnamefont {R.~L.}\ \bibnamefont {Leibel}},  \emph {et~al.},\
  }\href@noop {} {\bibfield  {journal} {\bibinfo  {journal} {Science}\ }\textbf
  {\bibinfo {volume} {341}} (\bibinfo {year} {2013})}\BibitemShut {NoStop}%
\bibitem [{\citenamefont {Hansen}\ \emph {et~al.}(2020)\citenamefont {Hansen},
  \citenamefont {Karslake}, \citenamefont {Woods}, \citenamefont {Read},\ and\
  \citenamefont {Wood}}]{hansen2020antibiotics}%
  \BibitemOpen
  \bibfield  {author} {\bibinfo {author} {\bibfnamefont {E.}~\bibnamefont
  {Hansen}}, \bibinfo {author} {\bibfnamefont {J.}~\bibnamefont {Karslake}},
  \bibinfo {author} {\bibfnamefont {R.~J.}\ \bibnamefont {Woods}}, \bibinfo
  {author} {\bibfnamefont {A.~F.}\ \bibnamefont {Read}}, \ and\ \bibinfo
  {author} {\bibfnamefont {K.~B.}\ \bibnamefont {Wood}},\ }\href@noop {}
  {\bibfield  {journal} {\bibinfo  {journal} {PLoS biology}\ }\textbf {\bibinfo
  {volume} {18}},\ \bibinfo {pages} {e3000713} (\bibinfo {year}
  {2020})}\BibitemShut {NoStop}%
\bibitem [{\citenamefont {Zhang}\ \emph {et~al.}(2011)\citenamefont {Zhang},
  \citenamefont {Lambert}, \citenamefont {Liao}, \citenamefont {Kim},
  \citenamefont {Robin}, \citenamefont {Tung}, \citenamefont {Pourmand},\ and\
  \citenamefont {Austin}}]{zhang2011acceleration}%
  \BibitemOpen
  \bibfield  {author} {\bibinfo {author} {\bibfnamefont {Q.}~\bibnamefont
  {Zhang}}, \bibinfo {author} {\bibfnamefont {G.}~\bibnamefont {Lambert}},
  \bibinfo {author} {\bibfnamefont {D.}~\bibnamefont {Liao}}, \bibinfo {author}
  {\bibfnamefont {H.}~\bibnamefont {Kim}}, \bibinfo {author} {\bibfnamefont
  {K.}~\bibnamefont {Robin}}, \bibinfo {author} {\bibfnamefont {C.-k.}\
  \bibnamefont {Tung}}, \bibinfo {author} {\bibfnamefont {N.}~\bibnamefont
  {Pourmand}}, \ and\ \bibinfo {author} {\bibfnamefont {R.~H.}\ \bibnamefont
  {Austin}},\ }\href@noop {} {\bibfield  {journal} {\bibinfo  {journal}
  {Science}\ }\textbf {\bibinfo {volume} {333}},\ \bibinfo {pages} {1764}
  (\bibinfo {year} {2011})}\BibitemShut {NoStop}%
\end{thebibliography}%

\end{document}